\documentclass[fleqn,usenatbib]{mnras}
\usepackage{newtxtext,newtxmath}
\usepackage[T1]{fontenc}
\DeclareRobustCommand{\VAN}[3]{#2}
\let\VANthebibliography\thebibliography
\def\thebibliography{\DeclareRobustCommand{\VAN}[3]{##3}\VANthebibliography}
\usepackage{color}
\usepackage{graphicx}	
\usepackage{amsmath}	

\title[Fine-structure transitions of C, N$^+$ and O$^{2+}$ with H]{Fine-structure transitions of the carbon
isoelectronic sequence C, N$^+$ and O$^{2+}$ induced by collisions with atomic hydrogen}

\author[P. G. Yan et al.]{
Pei-Gen Yan,$^{1}$\thanks{E-mail: peigen.yan@cfa.harvard.edu}
James F. Babb,$^{1}$\thanks{E-mail: jbabb@cfa.harvard.edu}
\\
$^{1}$Center for Astrophysics \textbar\ Harvard \& Smithsonian, MS 14, 60 Garden St., Cambridge, MA 02138
}
\pubyear{2022}

\begin{document}
\label{firstpage}
\pagerange{\pageref{firstpage}--\pageref{lastpage}}
\maketitle

\begin{abstract}
Fine-structure transitions can be involved in various processes including photon absorption, charge transfer and inelastic collision between ions, electrons and neutral atoms. 
We present fine-structure excitation and relaxation cross sections for
the collisions of the first few members of the carbon
isoelectronic sequence (C, N$^+$ and O$^{2+}$) with atomic hydrogen calculated
using quantum-mechanical methods.
For C, the scattering theory
and computational approach is verified by
comparison with previous calculations.
The rate coefficients for the collisional processes
are obtained.
For N$^+$ and O$^{2+}$, the transitions 
correspond to the lines [\ion{O}{iii}]
52~{\textmu}m, [\ion{O}{iii}] 88~{\textmu}m,
[\ion{N}{ii}] 122~{\textmu}m, and [\ion{N}{ii}] 205~{\textmu}m,
observed in the far-infrared in the local Universe and more recently
in high-redshift galaxies using radio interferometry.
The influence of different potentials on the cross sections and rate coefficients are demonstrated.  
\end{abstract}


\section{Introduction} \label{sec:intro}

In many astrophysical
environments, fine-structure levels of atoms and ions are populated by collisions
with electrons, hydrogen atoms, and hydrogen molecules, as well as by photons.
Excitation of an atom by such collisions might be followed by spontaneous emission leading to a
loss
of kinetic energy or ``cooling.''
The competition between relaxation of the fine-structure levels by collisions
and spontaneous emission affects the diagnostic potential of the observed lines.
Cross sections and rate coefficients for collisional relaxation can
be calculated and used in sophisticated modeling codes to interpret
astronomical observations.

In the far-infrared, the lines [\ion{O}{i}] 63~{\textmu}m, 
[\ion{O}{i}] 146~{\textmu}m, and 
[\ion{C}{ii}] 158~{\textmu}m are commonly
observed and utilized as probes of physical conditions;
accordingly the collisions of O and C$^+$ with electrons,
hydrogen atoms, and hydrogen molecules are well-studied.
Recently, the rest-frame far-infrared lines [\ion{O}{iii}] 52~{\textmu}m, [\ion{O}{iii}] 88~{\textmu}m,
[\ion{N}{ii}] 122~{\textmu}m, and [\ion{N}{ii}] 205~{\textmu}m,
became of interest\footnote{For completeness,
we note that [\ion{N}{iii}] 57~{\textmu}m is of interest,
but we do not treat it in this paper.} in observations of high-redshift
galaxies ($z>6$) using ALMA,
see, for example, \citet{Inoue2014}, \citet{Sugahara2021},
and \citet{RamosPadilla2022}.
While the O${}^{2+}$ and N${}^+$ lines above generally trace H\textsc{ii} regions and thus are expected to be most affected by electron collisions---for which
collision strengths are available,
see~\citet{tayal_electron_2011,tayal_transition_2017}---complete interpretations of these nascent high-$z$ observations are still in their infancy
and as far as we know, 
no data on the collisional cross sections
and rate coefficients for fine-structure excitation or de-excitation
of N$^+$ or O$^{2+}$ in collisions with H are available should
they be needed for modeling
or simulation.
In the present paper,
we treat the collisions 
\begin{equation}\label{OIIIH}
\textrm{A}(^3P_j)+ \textrm{H}(^2S_{1/2})\rightarrow \textrm{A}(^3P_{j'})+ 
\textrm{H}(^2S_{1/2})\,,
\end{equation}
where A can be C, N$^+$ and O$^{2+}$ and $(j,j')$ can be $(1,2)$, $(0,1)$, or $(0,2)$.
For ground state carbon atoms, quantum-mechanical calculations of
cross sections for
process~(\ref{OIIIH}) were previously reported by \citet{lr77}
and \citet{A07}.
Although the atomic structure
of the ions is in some ways similar to neutral carbon
we find that 
in collisions with hydrogen
atoms the resulting cross sections
and rate coefficients have quantitative differences.

\section{Theoretical Models}\label{TM}

We use a close-coupling scattering formulation~\citep{Mies1973a,lr77} with
the 
wave function $\Phi^{JM}$ expanded in the space-fixed basis $|j_aj_bj_{ab}lJM\rangle$ as 
\begin{equation}\label{wave}
\Phi^{JM}=\sum_{j_{ab},l} F_{j_{ab}l}^{JM}(R) |j_aj_bj_{ab}lJM\rangle\,,
\end{equation}
where
the two colliding systems are
labeled $a$ and $b$,
$F_{j_{ab}l}^{JM}(R)$ is the reduced radial wave function,
$R$ is the internuclear distance, $j_{ab}=j_a+j_b$ is the total electric angular
momentum, and $J=j_{ab}+l$ is the total angular momentum. 
With the total wave function (Eq.~\ref{wave}), the Schr\"{o}dinger equation that describes the internal motions of the C, N$^+$ or O$^{2+}$ ion and H system can be reduced to a set of coupled differential equations, expressed in atomic units (a.u.) as 
\begin{equation}\label{Sch}
\bigg[\frac{d}{dR^2}-\frac{l(l+1)}{R^2}+k_{j_a}^2\bigg]F_{j_{ab}l}^{J}(R)=2\mu \sum_{j'_aj'_{ab}l'}U^J_{j_aj_{ab}l;j'_aj'_{ab}l'} F_{j_{ab}l}^{J}(R) \,,
\end{equation}
with the wave number $k_{j_a}$ defined by $k^2_{j_a}=2\mu(E-\varepsilon_{j_a})$,
where $\mu$ is the reduced mass
of systems $a$ and $b$, $E$ is the total collision energy,
and $\varepsilon_{j_a}$ is
the fine-structure state splitting energy of the C, N$^+$, or O$^{2+}$ atom. 
We adopt
the values of $\varepsilon_{j_a}$
from NIST~\citep{NIST_ASD}:
For the C atom
\begin{equation}\label{FinE_C}
\varepsilon_0=0, \qquad \varepsilon_1=16.417\;\mathrm{cm}^{-1}, \qquad \varepsilon_2=43.413\;\mathrm{cm}^{-1} \,,
\end{equation} 
for the N$^+$ ion, 
\begin{equation}\label{FinE_N}
\varepsilon_0=0, \qquad \varepsilon_1=48.7\;\mathrm{cm}^{-1}, \qquad \varepsilon_2=130.8\;\mathrm{cm}^{-1} \,.
\end{equation}
and 
for the O$^{2+}$ ion, 
\begin{equation}\label{FinE_O}
\varepsilon_0=0, \qquad \varepsilon_1=113.178\;\mathrm{cm}^{-1}, \qquad \varepsilon_2=306.174\; \mathrm{cm}^{-1} \,.
\end{equation}
$U$ is the coupling matrix given by the summation of the $V_\mathrm{el}$ and $V_\mathrm{so}$ matrices in the basis $|j_aj_bj_{ab}lJM\rangle$. The
$V_\mathrm{so}$ matrix is diagonal and the nonzero elements are the fine-structure splittings that are given above in
(\ref{FinE_C}), (\ref{FinE_N}),
or (\ref{FinE_O}). For the elements of the potential energy $V_\mathrm{el}$ matrix, we adopt the expression described in \cite{lr77,flower_molcol:_2000,KJD06}. Then the cross sections for fine-structure transitions are calculated by
\begin{equation}
\sigma_{j_a \rightarrow j'_a}
(E) =\sum_{J} \sigma_{j_a \rightarrow j'_a}^J
(E)\,, 
\end{equation}
\begin{equation}
 \sigma_{j_a \rightarrow j'_a}^J
 (E)=\frac{\pi}{k_{j_a}^2}\frac{2J+1}{(2j_a+1)(2j_b+1)}
\sum_{j_{ab}lj'_{ab}l'}|T^J_{j'_aj'_{ab}l';j_aj_{ab}l}|^2\,, \label{cross}
\end{equation}
where $\sigma_{j_a \rightarrow j'_a}^J$ are the partial cross sections and the $T$ matrix is defined by $T^J=-2i K^J(I-iK^J)^{-1}$, where $K^J$ is the open channel reaction matrix defined in \citet{johnson_multichannel_1973}. 
The maximum value for the partial waves $J$ is set as 1000
and 
for the maximum collision energy used
(9.05~eV), the partial cross section converged
for CH at about 300, for NH$^+$ at about 500,
and for OH$^{2+}$ at about 990.

\section{Interatomic Potentials}\label{pot}

\begin{figure}
	\includegraphics[width=\columnwidth]{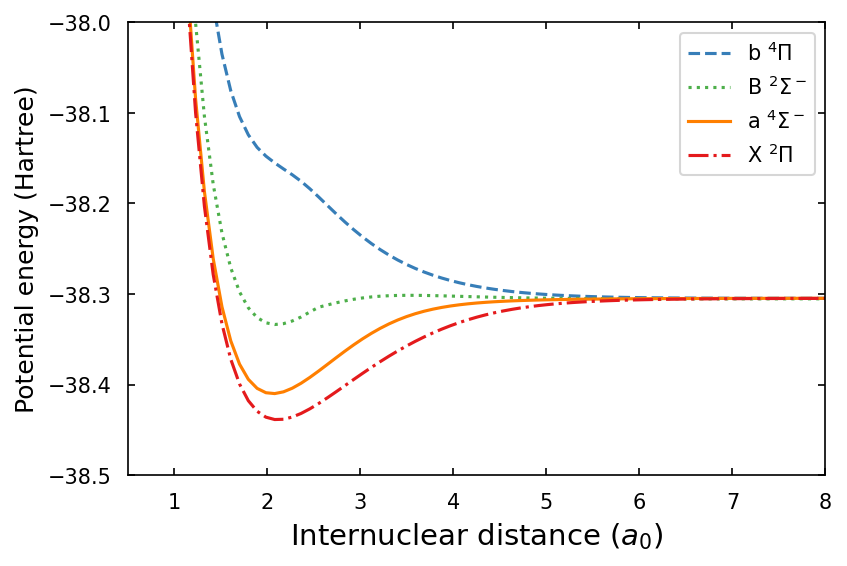}
    \caption{Potential energies (in Hartrees) for the CH system as functions of the internuclear distance $R$ (in $a_0$) labeled
    from the top down at the equilibrium distance
    of the lowest state: b$^4\Pi$ (blue dashed line), B$^2\Sigma^-$ (green dotted line), a$^4\Sigma^-$ (orange solid line) and X$^2\Pi$ (red dashdot line).}
    \label{fig:Cpot}
\end{figure}

For the CH system,  the four electronic states involved in the fine-structure transitions of C($^3$P) in collisions with H($^2$S) are the b$^4\Pi$, B$^2\Sigma^-$,
a$^4\Sigma^-$, and X$^2\Pi$ states.
The present calculations, using the 
MRCI-DKH (Douglas-Kroll-Hess) method with the \texttt{aug-cc-pV5Z-dk} basis within \textsc{Molpro} 2010.1~\citep{MOLPRO_brief}, are shown in Fig.~\ref{fig:Cpot}.
The calculated potential energies were smoothly joined to
the long-range potential function $-C_6/R^6$ 
at $R=9\,a_0$, where we used the value $C_6=20.3$ a.u. \citep{yau_fine-structure_1976}.   

\begin{figure}
	\includegraphics[width=\columnwidth]{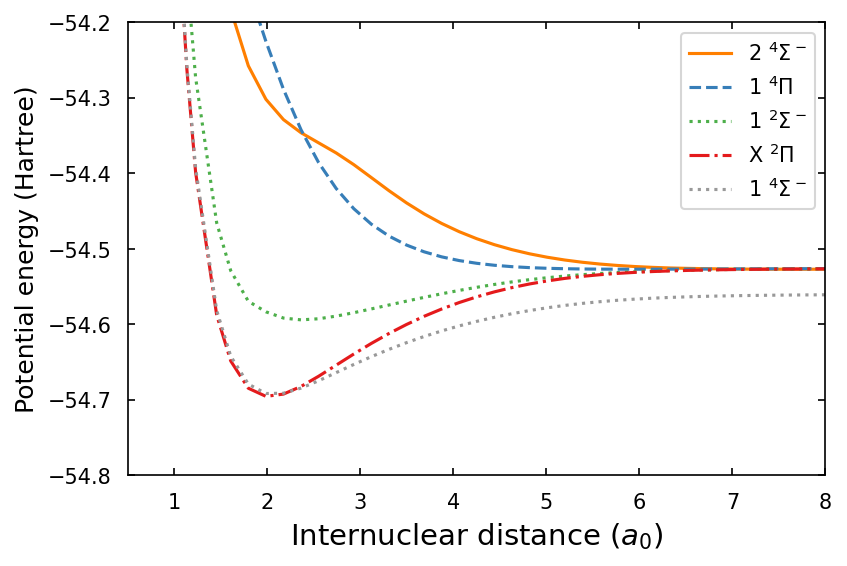}
    \caption{Potential energies for the NH$^+$ system as  functions of the internuclear distance $R$. The 1$^4\Sigma^-$ state 
    correlating to N($^4$S)--H$^+$($^1$S)
    is labeled with the gray dotted line. 
    The other states correlate
    to N$^+$($^3$P)--H($^2$S) and are labeled from the top down at
    the equilibrium distance of the lowest state: 1$^4\Pi$ (blue dashed line), 2$^4\Sigma^-$ (orange solid line),  1$^2\Sigma^-$ (green dotted line) and X$^2\Pi$ (red dashdot line).} 
    \label{fig:N+pot}
\end{figure}

For the NH$^+$ system,
the four electronic states
involved in the fine-structure transitions of
N$^+$($^3$P) in collision with H($^2$S) are
the 2$^4\Sigma^-$, 1$^4\Pi$, 1$^2\Sigma^-$, 
and X$^2\Pi$ states.
The present calculations, using the 
MRCI-DKH (Douglas-Kroll-Hess) method with 
the \texttt{aug-cc-pV5Z-dk} basis within \textsc{Molpro} 2010.1~\citep{MOLPRO_brief}, are shown in 
Fig.~\ref{fig:N+pot}, 
along with the 1$^4\Sigma^-$ state correlating to 
N($^4$S) + H$^+$($^1$S).
The potentials were joined at $R=17\,a_0$ to the long-range charge-induced dipole interaction $-\alpha_\mathrm{H}/2R^4$ between N$^+$($^3$P) and H($^2$S), where $\alpha_\mathrm{H} =4.5$ a.u. is the polarizability of the hydrogen atom. 

\begin{figure}
	\includegraphics[width=\columnwidth]{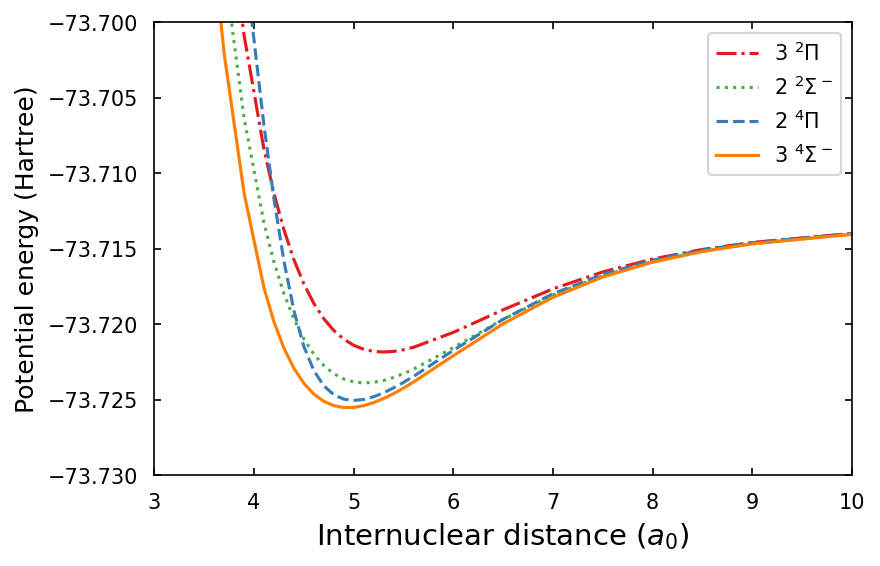}
    \caption{Potential energies for the OH$^{2+}$ system as functions of the internuclear distance $R$,
    labeled from the top down at the equilibrium distance
    of the lowest state:
    3$^2\Pi$ (red dashdot line), 2$^2\Sigma^-$ (green dotted line), 2$^4\Pi$ (blue dashed line), and 3$^4\Sigma^-$ (orange solid line) from \citet{de_melo_exploring_2021}.  }
    \label{fig:O2+pot}
\end{figure}

For the OH$^{2+}$ system, the four electronic states that are involved in the fine-structure transitions of O$^{2+}$($^3$P) in collision with H($^2$S) are the  3$^2\Pi$, 2$^2\Sigma^-$, 2$^4\Pi$, and 3$^4\Sigma^-$ states.
We adopted the recent MRCI+Q calculations using the \texttt{aug-cc-pV5Z} basis that were presented by \citet{de_melo_exploring_2021} and which
we show in Fig.~\ref{fig:O2+pot}.
The long-range charge-induced dipole interaction
potential $-\alpha_\mathrm{H}/R^4$ between O$^{2+}$($^3$P) ion and the H($^2$S) atom was smoothly connected to the potentials
at $R = 20\,a_0$.  

\section{Scattering calculations}\label{SC}

Using the theoretical quantum description of a collision between two open-shell atoms in arbitrary angular momentum states derived by \citet{KJD06},
the scattering calculations were performed using a Python code
that we wrote based partially on the \textsc{Molcol} Fortran~77 code \citep{molcol2019}
utilizing Johnson’s
multichannel logarithmic derivative algorithm \citep{johnson_multichannel_1973}. 
In order to prove the reliability of the present approach, the Python code was tested by calculating cross sections
for the C$^{+}$-H, the Si$^{+}$-H, and the O-H systems \citep{yan_babb_fine-structure_2022}; in all these cases, 
the corresponding
rate coefficients were found to be in very good agreement with those of preceding calculations
from the literature when we used
the potential energy data of the original sources. In the following, we will also show that our calculations of the cross sections and rate coefficients for the C-H system 
are in good accord with those of others. 

Having established the reliability of our theory and code, 
the cross sections were computed for collisional energies up to about 8.6~eV. 
The scattering equations were integrated from $R=1\,a_0$ to $R=30 \,a_0$.
The corresponding excitation and relaxation rate coefficients 
were computed for temperatures up to 10,000~K by averaging over a
Maxwellian energy distribution,
\begin{equation}\label{rate-coeff}
k_{j_a \rightarrow j'_a}(T)=\bigg(\frac{8}{\pi \mu k_B^3T^3}\bigg)^{1/2}\int_0^{\infty} \sigma_{j_a \rightarrow j'_a} (E_k) e^{-E_k/k_{B}T}E_k dE_k \,,
\end{equation}
where $T$ is the temperature,
$k_B$ is Boltzmann's constant, $\sigma$ is the cross section,
and $E_k$ is the kinetic energy.

\section{Results}\label{RES}

\begin{figure}
	\includegraphics[width=\columnwidth]{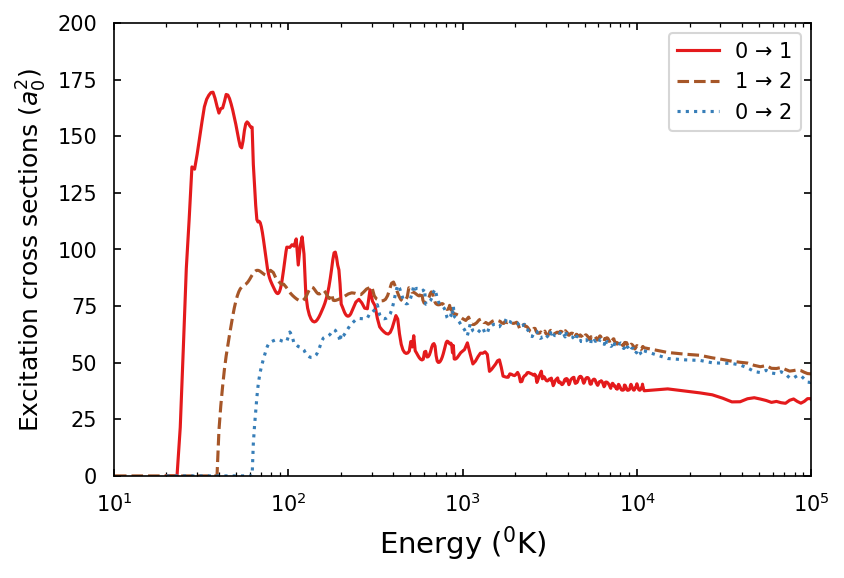}
	\includegraphics[width=\columnwidth]{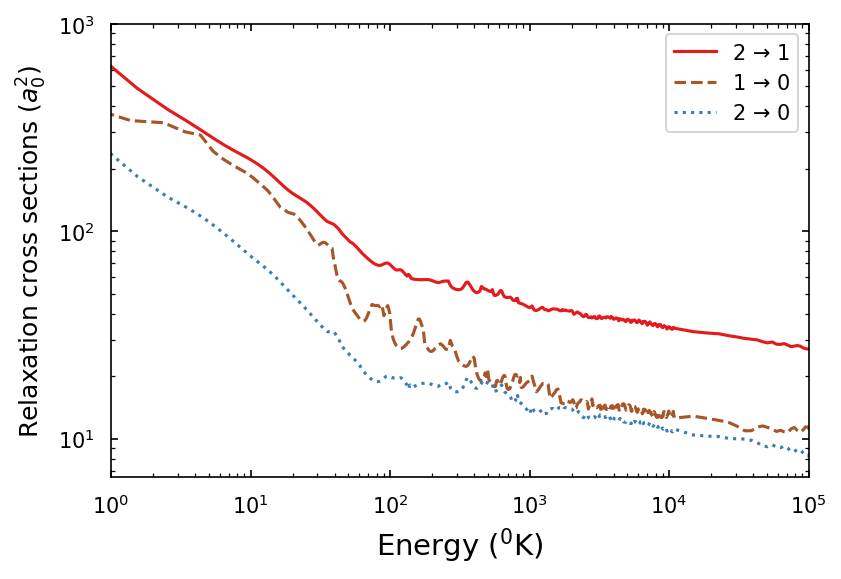}
    \caption{Excitation and relaxation cross sections for fine-structure transitions of C in collision with H. The three transitions $(j\rightarrow j')$ are labeled with different colors. }
    \label{fig:CExsct}
\end{figure}

\begin{figure}
	\includegraphics[width=\columnwidth]{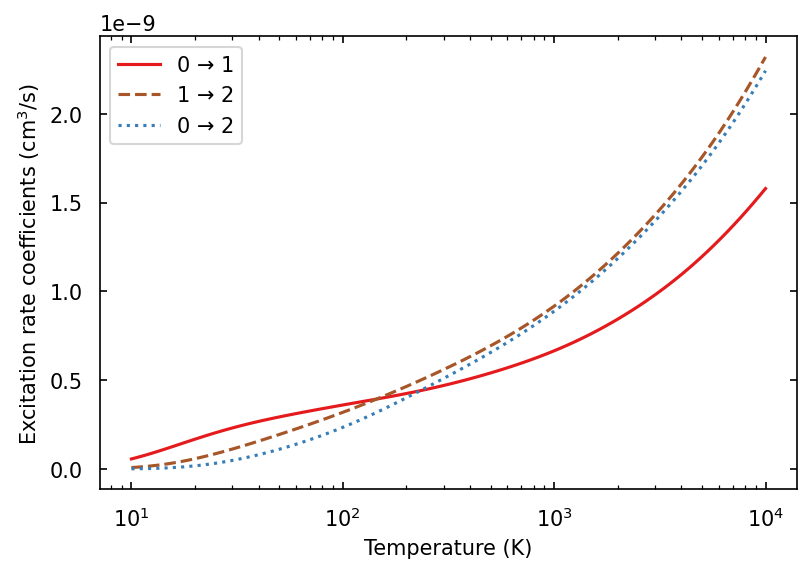}
	\includegraphics[width=\columnwidth]{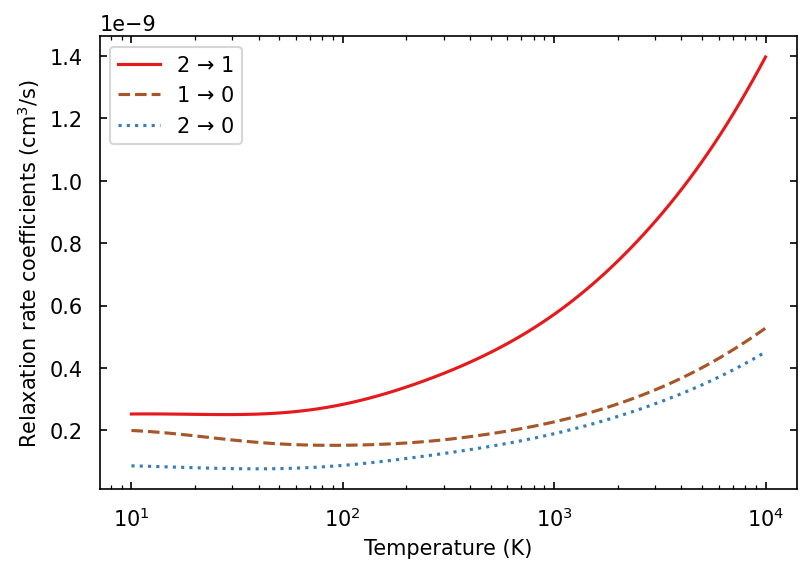}	
    \caption{Excitation and relaxation rate coefficients for fine-structure transitions of C in collision with H. The three transitions $(j\rightarrow j')$ are labeled with different colors. }
    \label{fig:CExrate}
\end{figure}

\begin{figure}
	\includegraphics[width=\columnwidth]{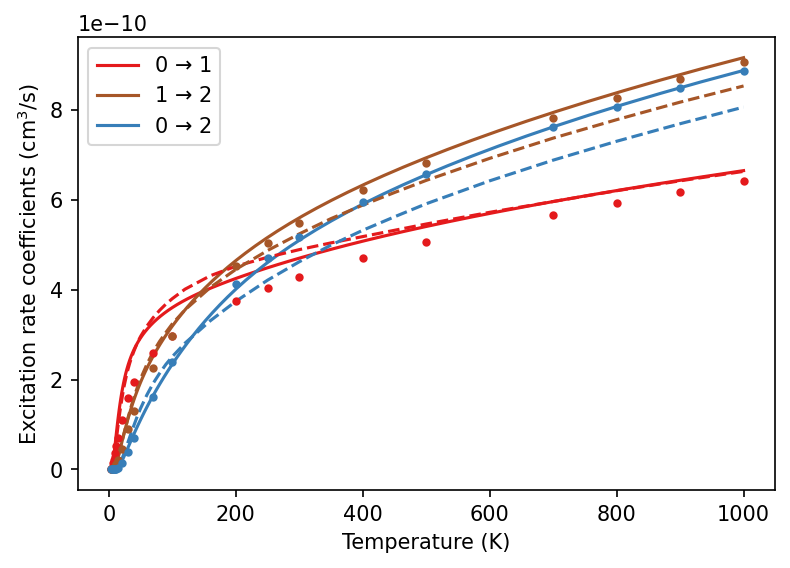}
    \caption{Comparison of temperature dependent excitation rate coefficients of the present calculations
     (solid lines)with those  of \citet{A07} 
    (large dots) and with those of \citet{lr77} (dashed lines).  }
    \label{fig:CHcom}
\end{figure}

With the CH potentials as described
above, the calculated 
excitation and relaxation cross 
sections \footnote{Note that the cross sections correspond
strictly to the process (\ref{OIIIH}) and we calculate
these to energies  $\sim 10^5$~K to have
a suitable integrand for the rate coefficient
of~Eq.~(\ref{rate-coeff}). In a gas 
of temperature $~\sim 10,000$~K or higher 
cross sections corresponding
to other channels such as $\textrm{C}(^1D)$ and $\textrm{C}(^1S)$
might be opened.}
are shown in Fig.~\ref{fig:CExsct}. 
The corresponding rate coefficients as functions of temperature are given in Fig.~\ref{fig:CExrate}
and Table~\ref{tab:chrate}.
\begin{table}
	\centering
	\caption{Rate coefficients (in units of 10$^{-9}$ cm$^3$s$^{-1}$) for the fine-structure excitation  and relaxation of C($^3P_j$) by H.}
	\label{tab:chrate}
	\begin{tabular}{lccccccccr} 
		\hline
         \multicolumn{1}{c}{$T$(K)} 
		  & $j=0 \rightarrow j'=1$ &  $j=0 \rightarrow j'=2$ &  $j=1 \rightarrow j'=2$\\
		\hline
		100 & 0.360 & 0.234 & 0.319\\
		200 & 0.425 & 0.401 & 0.465\\
		500 & 0.540 & 0.656 & 0.693\\
		700 & 0.596 & 0.762 & 0.794\\
		1000 & 0.665 & 0.888 & 0.917\\
		2000 & 0.844 & 1.185 & 1.215\\
		5000 & 1.195 & 1.707 & 1.755\\
		7000 & 1.368 & 1.948 & 2.010\\
		10000 & 1.579 & 2.242 & 2.320\\
		\hline
		\multicolumn{1}{c}{$T$(K)} & $j=1 \rightarrow j'=0$ &  $j=2 \rightarrow j'=0$ & $ j=2 \rightarrow j'=1$\\
		\hline
	    10 & 0.199 & 0.086 & 0.252\\
	    20 & 0.182 & 0.080 & 0.251\\
	    50 & 0.157 & 0.077 & 0.256\\
    	70 & 0.153 & 0.081  & 0.265\\
		100 & 0.152 & 0.088 & 0.283\\
		200 & 0.159 & 0.110 & 0.339\\
		500 & 0.189 & 0.149 & 0.450\\
		700 & 0.205 & 0.167 & 0.504\\
		1000 & 0.227 & 0.189 & 0.572\\
		2000 & 0.284 & 0.245 & 0.743\\
		5000 & 0.400 & 0.346 & 1.061\\
		7000 & 0.458 & 0.393 & 1.213\\
		10000 & 0.527 & 0.451 & 1.397\\
		\hline
	\end{tabular}
\end{table}
From Fig.~\ref{fig:CExrate}, we see that at low temperatures (or relatively small kinetic energies), the excitation rate coefficient of the $0 \rightarrow 1$ transition is larger than those of the $1 \rightarrow 2$ and $0 \rightarrow 2$ transitions. The energy differences of the $0 \rightarrow 1$ transition are the smallest leading to this being the strongest transition. With increasing temperature (kinetic energy), we find that the rate coefficients for the $0 \rightarrow 1$ transition drop to being the smallest, such behavior is explained by the ``forbidden'' selection rules demonstrated by \citet{monteiro_excitation_1987}.
This ``forbidden'' selection rule may explain the lower relaxation rate coefficients of $2 \rightarrow 0$ and $1 \rightarrow 0$ transitions compared to that of the $2 \rightarrow 1$ transition. In order to further prove the reliability of the present approach, we also made
comparisons of the present calculations with those of \citet{lr77}
and of \citet{A07}, as shown in Fig.~\ref{fig:CHcom}.
It can be seen that the present calculations are in good agreement
with the other works,
especially for the $0 \rightarrow 1$ transitions. The discrepancies for the rate coefficients of the $1 \rightarrow 2$ and $0 \rightarrow 2$ transitions could be caused by the different potentials we used. 
Comparing Figs.~\ref{fig:CExsct} and~\ref{fig:CExrate}, 
it is apparent that the relaxation cross sections decrease as a function of energy, while the relaxation rate coefficients increase as a function of temperature, a consequence
of the energy dependence in the integrand of 
Eq.~(\ref{rate-coeff}).
For example, the log-log plot of Fig.~\ref{fig:CExsct} 
shows that the relaxation cross sections decrease slowly
for energy $E_k > 1000\,^0\textrm{K}$
and 
thus the relaxation rate coefficients shown in Fig.~\ref{fig:CExrate} increase for $T > 100\,\textrm{K}$.    

\begin{figure}
	\includegraphics[width=\columnwidth]{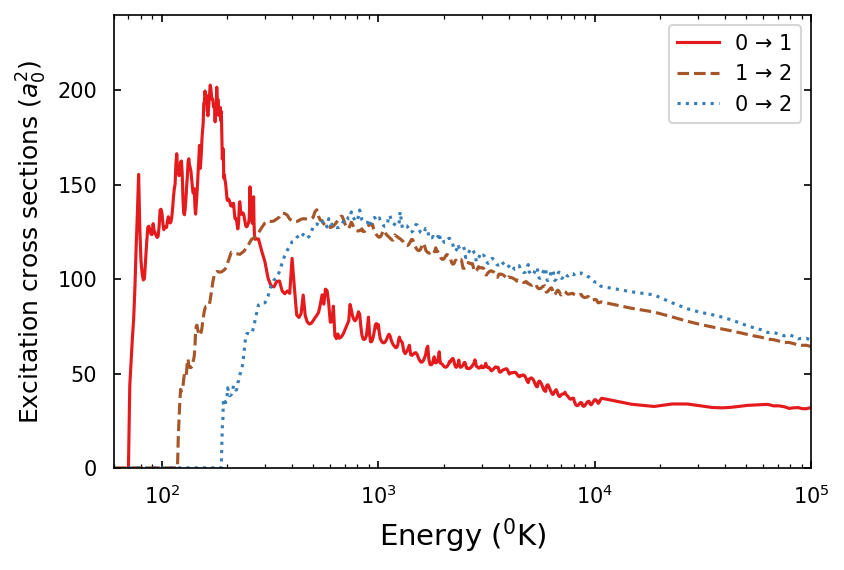}
    \includegraphics[width=\columnwidth]{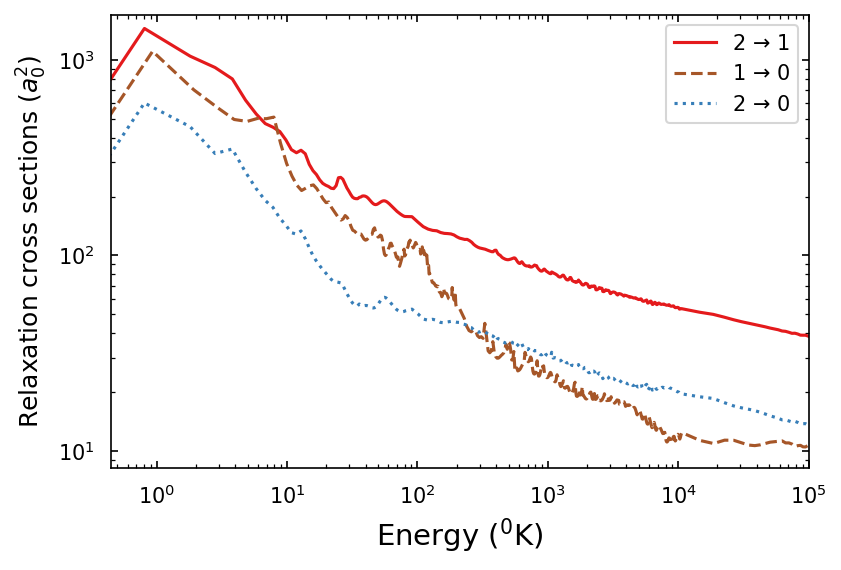}
    \caption{Excitation and relaxation cross sections for fine-structure transitions of N$^+$ in collision with H. The three transitions $(j\rightarrow j')$ are labeled with different colors.}
    \label{fig:N+sct}
\end{figure}

\begin{figure}
    \includegraphics[width=\columnwidth]{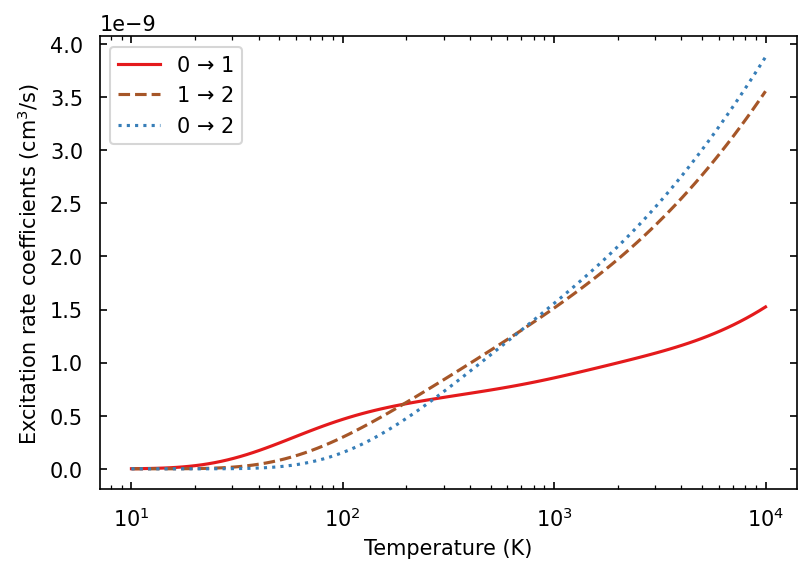}
    \includegraphics[width=\columnwidth]{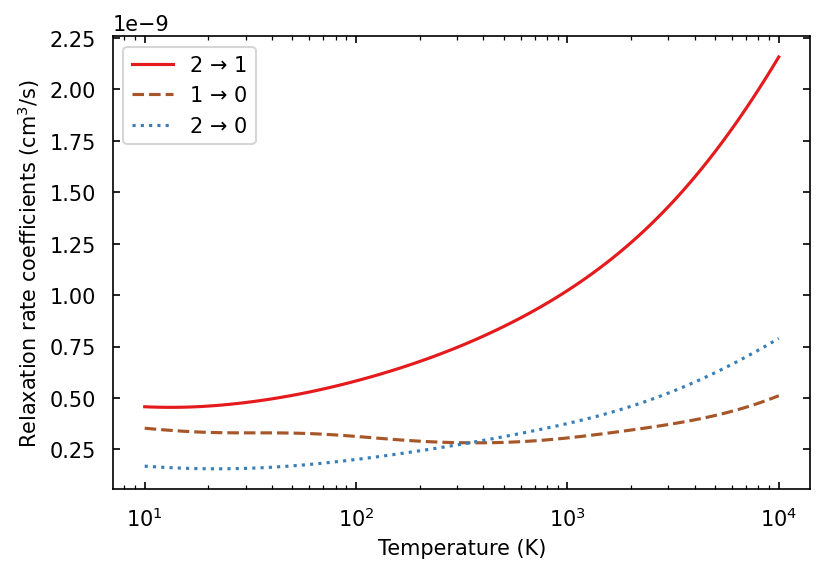}
    \caption{Excitation and relaxation rate coefficients for fine-structure transitions of $\textrm{N}^+$ in collision with H. The three transitions $(j\rightarrow j')$ are labeled with different colors.}
    \label{fig:N+rate}
\end{figure}

With the NH$^+$ potentials as described
above, the calculated 
excitation and relaxation cross sections 
are shown in Fig.~\ref{fig:N+sct}.
The corresponding rate coefficients for the fine structure transitions of N$^+$ in collision with H as a function of
temperature are given in Fig.~\ref{fig:N+rate}
and Table~\ref{tab:nh+rate}.
\begin{table}
	\centering
	\caption{Rate coefficients (in units of 10$^{-9}$ cm$^3$s$^{-1}$) for the fine-structure excitation and relaxation of N$^+$($^3P_j$) by H.}
	\label{tab:nh+rate}
	\begin{tabular}{lccccccccr} 
		\hline
         \multicolumn{1}{c}{$T$(K)} 
		  &$ j=0 \rightarrow j'=1$ & $ j=0 \rightarrow j'=2$ &  $j=1 \rightarrow j'=2$\\
		\hline
		100 & 0.466 & 0.153 & 0.298\\
		200 & 0.613 & 0.475 & 0.626\\
		500 & 0.742 & 1.072 & 1.116\\
		700 & 0.793 & 1.306 & 1.305\\
		1000 & 0.856 & 1.558 & 1.514\\
		2000 & 0.998 & 2.093 & 1.973\\
		5000 & 1.228 & 3.002 & 2.764\\
		7000 & 1.353 & 3.408 & 3.126\\
		10000 & 1.524 & 3.878 & 3.556\\
		\hline
		\multicolumn{1}{c}{$T$(K)} & $j=1 \rightarrow j'=0 $&  $j=2 \rightarrow j'=0$ & $ j=2 \rightarrow j'=1$\\
		\hline
        10 & 0.353 & 0.169 & 0.458\\
	    20 & 0.333 & 0.156 & 0.461\\
        50 & 0.330 & 0.171 & 0.513\\
        70 & 0.324 & 0.184  & 0.544\\
		100 & 0.314 & 0.201 & 0.583\\
		200 & 0.290 & 0.244 & 0.678\\
		500 & 0.285 & 0.313 & 0.848\\
		700 & 0.292 & 0.342 & 0.927\\
		1000 & 0.306 & 0.376 & 1.022\\
		2000 & 0.345 & 0.460 & 1.256\\
		5000 & 0.415 & 0.624 & 1.698\\
		7000 & 0.455 & 0.700 & 1.907\\
		10000 & 0.511 & 0.790 & 2.158\\
		\hline
	\end{tabular}
\end{table}
Compared to the CH system, the cross sections and rate coefficients of the $1 \rightarrow 2$ and $0 \rightarrow 2$ transitions for the NH$^+$ system are larger because of the stronger long-range attractions and deeper potential wells as shown in Fig.~\ref{fig:N+pot}.
Furthermore, we find that for NH$^+$ the $1 \rightarrow 2$ transition rate coefficient is the largest 
compared to CH where the $0 \rightarrow 2$ transition
is the largest. This phenomenon might be caused by the different shapes of the potentials for the two systems: The 1$^4\Sigma^-$ state of the CH system has a much deeper well than the 2$^4\Sigma^-$ state of the NH$^+$ system. Similarly to the CH system, 
the excitation rate coefficients of the $0 \rightarrow 1$ transition are also the smallest among the three transitions because of the ``forbidden'' selection rule.  
$\textrm{N}^+$ can be removed
by the exothermic charge transfer process~\citep{Butler1979}
\begin{equation}\label{rate-exo}
\textrm{N}^+(^3P) + \textrm{H}(^2S) \rightarrow \textrm{N}(^4S^\mathrm{o} )+ \textrm{H}^+ + 0.9467~\textrm{eV}.
\end{equation}
The rate coefficients were calculated by \citet{Lin2005}
and they decrease from 
a value of about $10^{-13}\,\textrm{cm}^3/\textrm{s}$
at $10,000\,\textrm{K}$
and appear to be 
considerably slower than the relaxation
process which is no less
than $10^{-10}\,\textrm{cm}^3/\textrm{s}$
even at $10\,\textrm{K}$.

\begin{figure}
	\includegraphics[width=\columnwidth]{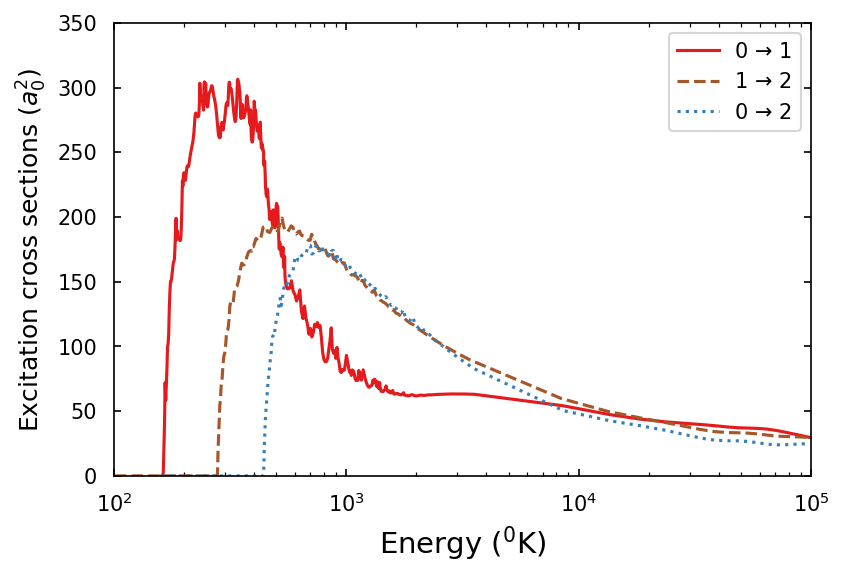}
	\includegraphics[width=\columnwidth]{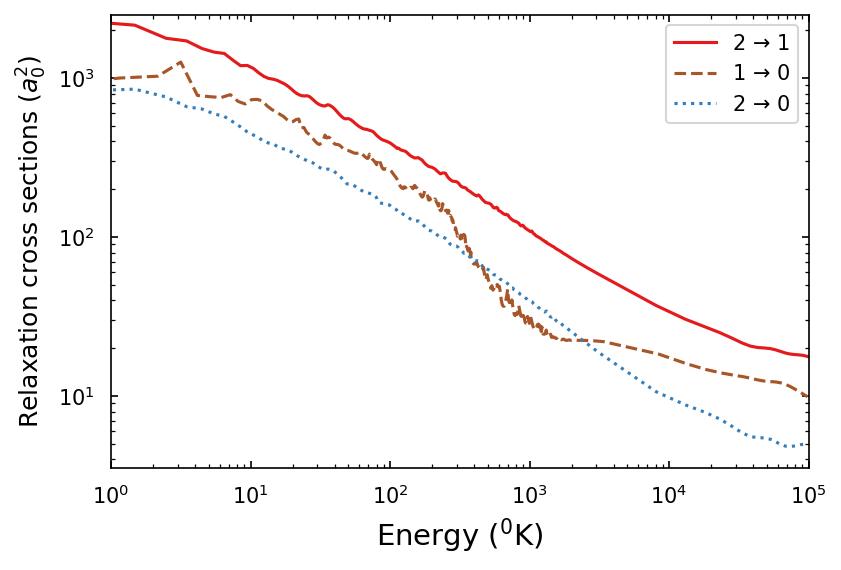}
    \caption{Excitation and relaxation cross sections for fine-structure transitions of O$^{2+}$ in collision with H. The three transitions $(j\rightarrow j')$ are labeled with different colors.}
    \label{fig:O2sct}
\end{figure}

\begin{figure}
	\includegraphics[width=\columnwidth]{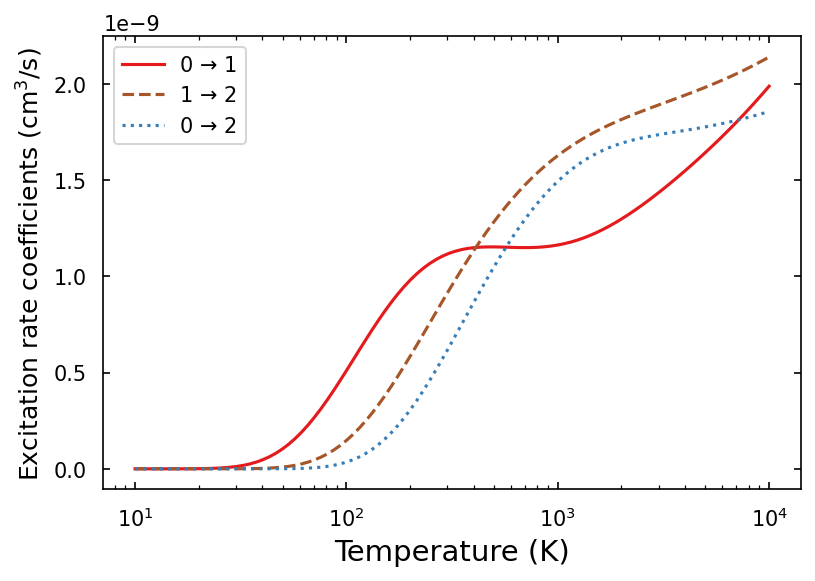}
    \includegraphics[width=\columnwidth]{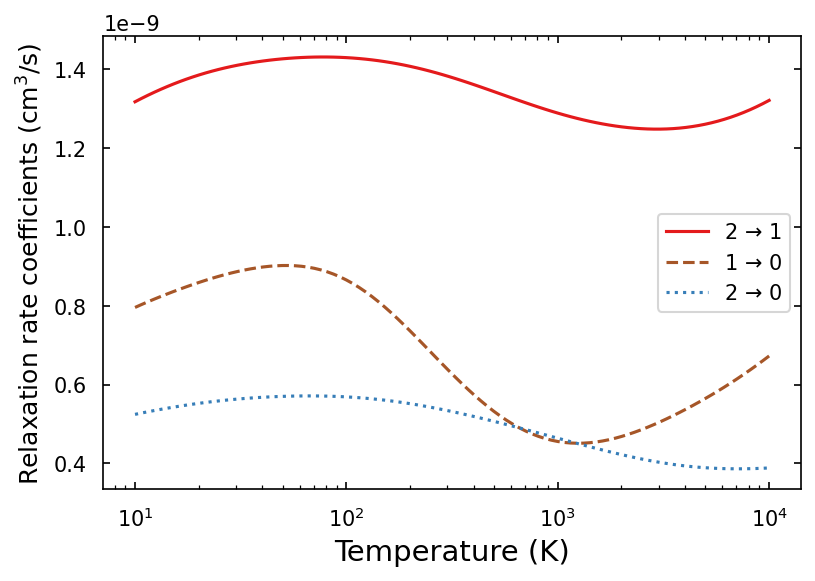}
    \caption{Excitation and relaxation rate coefficients for fine-structure transitions of O$^{2+}$ in collision with H. The three transitions $(j\rightarrow j')$ are labeled with different colors.}
    \label{fig:O2rate}
\end{figure}

With the OH$^{2+}$ potentials as described
above, the calculated cross sections 
for fine structure transition of O$^{2+}$ in collision with H are shown in Fig.~\ref{fig:O2sct}.
The corresponding rate coefficients are given in Fig.~\ref{fig:O2rate}
and Table~\ref{tab:oh2+rate}.
\begin{table}
	\centering
	\caption{Rate coefficients  (in units of 10$^{-9}$ cm$^3$s$^{-1}$) for the fine-structure excitation and relaxation of O$^{2+}$($^3P_j$) by H.}
	\label{tab:oh2+rate}
	\begin{tabular}{lccccccccr} 
		\hline
         \multicolumn{1}{c}{$T$(K)} 
		  & $j=0 \rightarrow j'=1$ &  $j=0 \rightarrow j'=2$ &  $j=1 \rightarrow j'=2$\\
		\hline
		100 & 0.510 & 0.348 & 0.147\\
		200 & 0.979 & 0.305 & 0.583\\
		500 & 1.152 & 1.050 & 1.284\\
		700 & 1.149 & 1.296 & 1.474\\
		1000 & 1.162 & 1.493 & 1.626\\
		2000 & 1.296 & 1.690 & 1.814\\
		5000 & 1.643 & 1.776 & 1.982\\
		7000 & 1.800 & 1.809 & 2.051\\
		10000 & 1.987 & 1.856 & 2.138\\
		\hline
		\multicolumn{1}{c}{$T$(K)} & $j=1 \rightarrow j'=0$ &  $j=2 \rightarrow j'=0$ &  $j=2 \rightarrow j'=1$\\
		\hline
		10 & 0.796 & 0.525 & 1.318\\
		20 & 0.859 & 0.552 & 1.386\\
		50 & 0.902 & 0.570 & 1.427\\
		70 & 0.896 & 0.571  & 1.431\\
		100 & 0.866 & 0.569 & 1.430\\
		200 & 0.737 & 0.552 & 1.408\\
		500 & 0.532 & 0.507 & 1.344\\
		700 & 0.483 & 0.486 & 1.316\\
		1000 & 0.456 & 0.464 & 1.288\\
		2000 & 0.469 & 0.421 & 1.249\\
		5000 & 0.566 & 0.387 & 1.255\\
		7000 & 0.614 & 0.384 & 1.279\\
		10000 & 0.673 & 0.387 & 1.318\\
		\hline
	\end{tabular}
\end{table}
From Figs.~\ref{fig:O2sct} and  \ref{fig:O2rate}, we see that at the lowest temperatures both the cross sections and rate coefficients of the OH$^{2+}$ system are larger than those of the CH and NH$^+$ systems. This is caused by the stronger long-range interaction $-\alpha_\mathrm{H}/R^4$ of the OH$^{2+}$ system compared to the other two systems. 
The four wells of the OH$^{2+}$ potentials,
see Fig.~\ref{fig:O2+pot},
lead to the mild increase of the  $0 \rightarrow 1$ transition for temperatures larger than 4000~K.
$\textrm{O}^{2+}$ can be removed by the charge transfer process~\citep{ButlerBender1979}
\begin{equation}\label{rate-cx-oxy}
\textrm{O}^{2+}(^3P) + \textrm{H}(^2S) \rightarrow \textrm{O}^+(^4P )+ \textrm{H}^+ + 6.701~\textrm{eV}.
\end{equation}
The rate coefficient is about $10^{-9}\,\textrm{cm}^3/\textrm{s}$ over
the temperature range $200<T<10,000\,\textrm{K}$ \citep{Honvault1995},
which is comparable to our calculated
fine-structure relaxation rate coefficients
shown in Table~\ref{tab:oh2+rate}.
This might be an important consideration for modeling applications. 

\begin{figure}
	\includegraphics[width=\columnwidth]{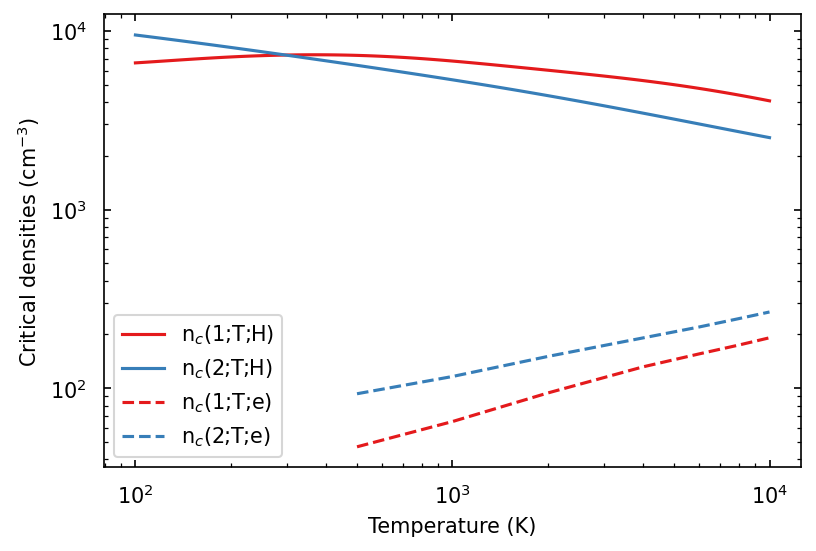}
    \caption{Critical densities for the relaxation of N$^{+}$ in collision with hydrogen (solid lines) and electrons (dashed lines). For electrons, data from~\citet{tayal_electron_2011} were used. } 
    \label{fig:Ndensity}
\end{figure}

\begin{figure}
	\includegraphics[width=\columnwidth]{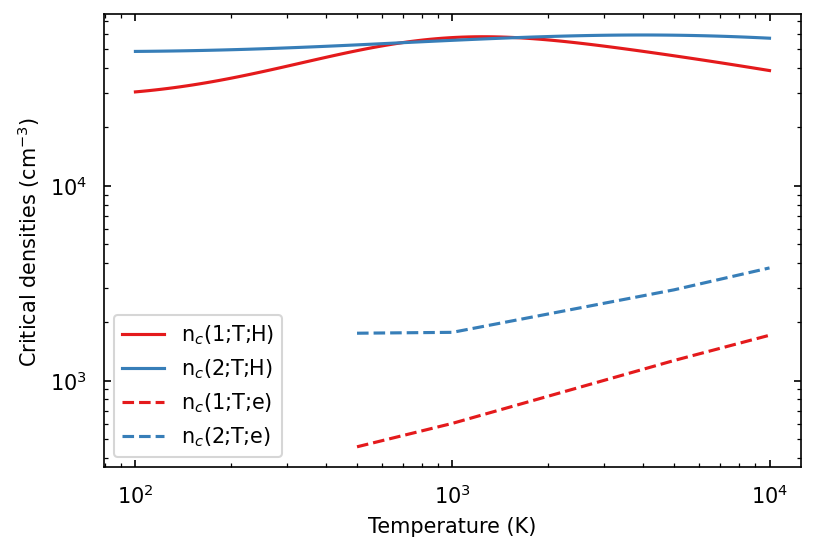}
    \caption{Critical densities for the  relaxation of O$^{2+}$ in collision with hydrogen (solid lines) and electrons (dashed lines). For electrons,
    data from~\citet{tayal_transition_2017} were used. }
    \label{fig:O2density}
\end{figure}

The critical density $n_c$, an important parameter to indicate whether collisions affect
the presence of fine structure lines,
is defined as
\begin{equation}
n_{c}(j;T;x)
=\frac{\sum_{j'} A(j \rightarrow j')}{\sum_{j'} k_{j \rightarrow j'}(T;x)} \,,
\end{equation}
where $A(j \rightarrow j')$ is the transition probability~\citep{mendoza_recent_1983}
and 
$k_{j \rightarrow j'}(T;x)$
is the relaxation rate coefficient for collisions with species $x$, which
maybe be hydrogen (H) or electrons (e).
Our calculated  critical densities 
for N$^+$ and O$^{2+}$, respectively, and H
are shown in Figs.~\ref{fig:Ndensity} 
and ~\ref{fig:O2density}.
We also plot the critical densities
for electrons (e), which were
determined using the effective collision strengths 
from~\citet{tayal_electron_2011} for N$^+$ 
and from ~\citet{tayal_transition_2017} for O$^{2+}$.
The graphical data in Figs.~\ref{fig:Ndensity} and ~\ref{fig:O2density}
might be useful for 
considerations of the relative
importance of hydrogen and electron densities
at specific temperatures.
For example,
we can see that the critical densities induced by electrons are 10 to 10$^2$ times smaller than those induced by hydrogen atoms due to the strong electron-ion interactions. 
\section{Conclusion}\label{Con}

We introduced a methodology that yields consistent sets of rate coefficients for a number of cases. Based partially on the \textsc{Molcol} Fortran code \citep{molcol2019}, we developed a new Python code utilizing Johnson’s
multichannel logarithmic derivative algorithm \citep{johnson_multichannel_1973}. 
We presented
for the first time a comparison
of the excitation and relaxation cross sections and rate coefficients for fine-structure transitions in collision of C, N$^+$, or O$^{2+}$ and atomic hydrogen. The calculations are based on accurate CH (present work), NH$^+$ (present work), and OH$^{2+}$ \citep{de_melo_exploring_2021} potentials. 
The present research may be useful in
applications of far-infrared lines to astrophysical diagnostics.

\section*{Data Availability}
The data points corresponding to the rate coefficients plotted in Figs.~\ref{fig:N+rate} and \ref{fig:O2rate} are given
in LAMDA format~\citep{moldata2005} at Figshare~\citep{YanBabbFig22b}.

\section*{Acknowledgements}
This work was supported by NASA APRA grant 80NSSC19K0698.

\bibliographystyle{mnras}
\bibliography{reference} 
\label{lastpage}
\end{document}